# Photon localisation and Bloch symmetry breaking in luminal gratings

E. Galiffi[1], M.G. Silveirinha[2], P. A. Huidobro[2], and J.B. Pendry[1]

[1]The Blackett Laboratory, Department of Physics,
Imperial College London, London, SW7 2AZ UK

[2]Instituto de Telecomunicações, Instituto Superior Técnico-University of Lisbon,
Avenida Rovisco Pais 1, Lisboa, 1049-001 Portugal.

*Abstract*

In gratings travelling at nearly the velocity of light a symmetry breaking transition is observed between free-flowing fluid-like Bloch waves observed at lower grating velocities and, at luminal velocities, condensed, localised states of light captured in each period of the grating and locked to its velocity. We introduce a new technique for calculating in this regime and use it to study the transition in detail shedding light on the critical exponents, and the periodic oscillations in transmitted intensity seen in the pre-transition regime.

Spurred on by experimental advances in rapid modulation both at optical [1,2], THz [3-5], and GHz [6] frequencies, in recent years interest has grown in systems where the parameters vary with time: in electromagnetic systems, the topic of the present paper, but also in acoustic systems [7]. In fact many of the concepts are quite general and apply to any wave motion whatever the system. One of the simplest static structures is the Bragg grating whose study has shown the way to photonic crystals and to a host of other electromagnetic devices. Translational symmetry permits the analysis of gratings in terms of Bloch waves and their dispersion is characterised by a Bloch wave vector which shows band gaps: ranges of frequency within which the wave vector is complex and where light is reflected from the structure.

Localisation is usually encountered in disordered systems where electron transport makes a transition from a diffusive regime which is well understood, to a localised regime which effectively cuts off conductivity. The transition is a complex one with no complete theoretical understanding of the localised regime and in particular of the critical exponent which is known from computer simulations [8] to be 1.571 but for which there is no fundamental derivation. However there are other forms of localisation, one of which we present here, that have a much simpler nature and are fully soluble. Their solution may shed light on the more complex problem.

We shall be concerned for the main part with a simple generalisation of a Bragg grating of the form,

$$\begin{aligned}\varepsilon\left(x-c_g t\right)&=\varepsilon_1\left[1+2\alpha_\varepsilon\cos\left(gx-\Omega t\right)\right]\\ \mu\left(x-c_g t\right)&=\mu_1\left[1+2\alpha_\mu\cos\left(gx-\Omega t\right)\right]\end{aligned}\tag{1}$$

moving with velocity $c_g=\Omega/g$. The velocity of light in the background medium is $c_l=c_0\big/\sqrt{\varepsilon_1\mu_1}$. We stress that material comprising the grating does not move, rather the local properties are modulated in the synchronised form given above. This allows the structure to move with any velocity, unrestricted by the speed of light. This model has been widely adopted in time dependent studies of 'space-time crystals' [9] and of non-reciprocal systems [10,11]. Closely related models have been used to study topological aspects of so-called time-crystals [12]. Although the modified grating lacks the time symmetric properties of the static case, it is parity-time (PT) symmetric and supports Bloch waves which have been the basis of previous studies. A sketch is given in fig. 1 showing that as $c_g \to c_l$ the forward travelling waves tend to degeneracy. Their close proximity inevitably leads to very strong interaction between the forward waves which is responsible for their pathological symmetry breaking behaviour in the luminal regime.

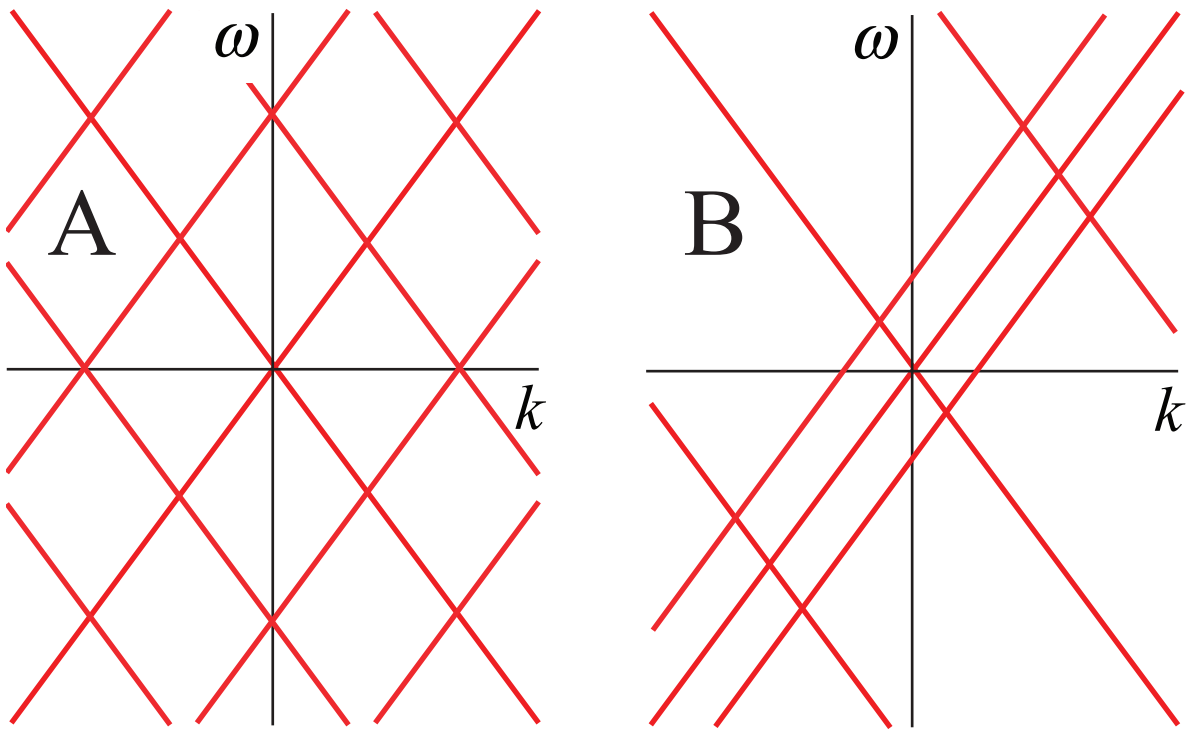

Fig. 1. A: Dispersion of waves in a stationary grating before interaction is taken account of. B: dispersion of waves in a moving grating. Note the asymmetry between forward and backward travelling waves in B brought about by the lack of time reversal symmetry and the close proximity of the forward waves which collapses to a degeneracy when $c_g = c_l$.

However, intriguingly and the motivation of this paper, there is a range of grating velocities within which the Bloch wave picture fails and no characterisation exists in terms of $\omega(k)$. There is a range of velocities, $c_- < c_g < c_+$, close to the speed of light within which light entering the medium is captured and localised inside each grating period where it is both amplified and compressed into an ever sharper pulse. In previous papers [13,14] we identified this behaviour as a phenomenon where amplification is associated with compression of the lines of force which are conserved.

Here we are concerned with the approach to localisation and the transition from Bloch-wave behaviour to formation of localised pulses. When $c_g < c_-$ (or $c_g > c_+$) light incident on the grating excites a mixture of Bloch waves leading to periodic oscillations of intensity within the grating caused by the grating alternately adding and subtracting energy from the waves with a period determined by the difference in speed of the Bloch waves, $c_{eff}$, and that of the grating. As $c_g \to c_-$ the Bloch wave velocity also approaches $c_-$ so that the period tends to infinity and at the same time the amplitude approaches infinity. In this way the oscillations presage the transition to localisation. In fact the period of these oscillations is related to exponential growth of energy density within the pathological region $c_- < c_g < c_+$. We calculate the period of these oscillations and the critical exponent with which it diverges at the transition. Within the pathological range the exponentially growing pulse will locate at a specific point within the grating period dictated by a subtle balance between its velocity keeping pace with the grating and the source of it's growing energy which is also a function of location within the grating.

The structure of our paper is as follows: first we introduce a new technique for solving the equations of motion based on the "method of trajectories". Not only does this enable efficient calculation of the grating properties, it also allows us to find analytic expressions for many of the quantities of interest. These we exploit in the next section to calculate critical exponents, pre-localisation oscillations and localisation of the pulse within the critical region. Our theory holds in any scenario where back-scattering is not significant.

*The equation of motion*

The transition to localisation is essentially an interaction between forward travelling waves which are highly degenerate near the transition. Therefore to study the transition in its purest form we choose a model in which there is no back scattering. This condition can be achieved exactly if $\varepsilon$ and $\mu$ are everywhere proportional,

$$\varepsilon\left(x-c_g t\right)=\mu\left(x-c_g t\right) \tag{2}$$

but is approximately true if the amplitude of the grating is sufficiently small. In the absence of back scattering the following relationships hold,

$$H_y=-\frac{1}{Z_m}E_z, \quad B_y=-Z_m D_z \tag{3}$$

where we have assumed a wave with electric fields aligned with the $z$ axis travelling in the forward direction along the $x$ axis. Here, $E_z, H_y$ are the electric and magnetic fields, and $D_z, B_y$ are the displacement vector and magnetic induction fields, respectively. We shall exploit this result to reduce Maxwell's equations to a partial differential equation (PDE) of the first order. Defining,

$$c_g=\Omega/g=\left(1+\delta\right)c_\ell \tag{4}$$

and working in a Gallilean frame co-moving with the grating. $X=x-c_g t$, we retrieve from Maxwell,

$$\frac{\partial}{\partial X}\left[c_l-c_g\right]D_z=-\frac{\partial}{\partial t}D_z, \tag{5}$$

A fuller description of this derivation is to had in [14]. Defining,

$$\psi=\left[c_l-c_g\right]D_z \tag{6}$$

gives the standard form,

$$\left(c_l-c_g\right)\frac{\partial\psi\left(X,t\right)}{\partial X}+\frac{\partial\psi\left(X,t\right)}{\partial t}=0 \tag{7}$$

which can be solved by the 'method of characteristics'. This approach seeks a characteristic trajectory, $X\left(t\right)$, along which $\psi\left(X,t\right)$ is constant. Then all that is necessary to solve the PDE is to trace the trajectory back to $t=0$ where the value of $\psi_0\left(X_0,t=0\right)$ gives the value of $\psi\left(X,t\right)$ at any subsequent time along the trajectory. To obtain the electric field we use (6) to give,

$$D_z\left[X\left(t\right)\right]=\frac{c_l\left[X_0\left(t=0\right)\right]-c_g}{c_l\left[X\left(t\right)\right]-c_g}D_z\left[X_0\left(t=0\right)\right] \tag{8}$$

The equation of motion along the trajectory is,

$$\frac{dX}{dt}=\left(c_l-c_g\right)=c_1\left(\frac{1}{1+2\alpha\cos\left(gX\right)}-\left(1+\delta\right)\right) \tag{9}$$

which can be solved for the time at which the trajectory reaches point $X$, by integration,

$$\int_{X_0}^{X} \frac{1}{c_l[X'] - c_g} dX' = t(X). \tag{10}$$

For a grating with a co-sinusoidal profile shown in (1), one finds

$$\begin{aligned}
\theta(X) &= t(X) - t_0 \\
&= \frac{-gX}{c_\ell g(1+\delta)} - \frac{1}{c_\ell g(1+\delta)^2} \frac{2}{\sqrt{\left(\frac{\delta}{1+\delta}\right)^2 - 4\alpha^2}} \tan^{-1}\left[\frac{\left(\frac{\delta}{1+\delta} - 2\alpha\right)\tan(gX/2)}{\sqrt{\left(\frac{\delta}{1+\delta}\right)^2 - 4\alpha^2}}\right], \\
&\left|\frac{\delta}{1+\delta}\right| > |2\alpha|, \\
&= \frac{-gX}{c_\ell g(1+\delta)} - \frac{1}{c_\ell g(1+\delta)^2} \frac{2}{\sqrt{4\alpha^2 - \left(\frac{\delta}{1+\delta}\right)^2}} \operatorname{ctnh}^{-1} \frac{\left(2\alpha - \frac{\delta}{1+\delta}\right)\tan(gX/2)}{\sqrt{4\alpha^2 - \left(\frac{\delta}{1+\delta}\right)^2}}, \\
&\left|\frac{\delta}{1+\delta}\right| < |2\alpha|, \quad \left|\left(2\alpha - \frac{\delta}{1+\delta}\right)\tan(gX/2)\right| > \sqrt{4\alpha^2 - \left(\frac{\delta}{1+\delta}\right)^2}
\end{aligned} \tag{11}$$

The trajectories shown in fig. 2 reveal the localisation transition. When the grating velocity is outside the critical points (panels A and C) these trajectories show a pulse of light advancing continuously through the grating structure, albeit in the backwards direction if $c_g > c_\ell$. Note also that the light pauses at certain positions within the grating: at $gX = 0$ for a sub luminal grating ( $\delta < 0$ ) where the local light velocity is a minimum, and at $gX = \pi$ for a superluminal grating ( $\delta > 0$ ) where the local light velocity is a maximum. In contrast for near luminal gratings with velocities inside the critical points (panels B and D) all trajectories show singularities which trap the light near that particular position. The singular points occur where the local velocity of light is the same as that of the grating and migrate to $gX = 0, \pi$ at either end of the critical zone, to coincide with the pause zones of fig. 2A and fig. 2C.

Curves in fig. 2C show how the extra luminal curves steepen at the origin for the subluminal case and at $gX/\pi = 1.0$ for the super luminal case. In fig. 2D the singularities migrate across the unit cell with increasing $\delta$.

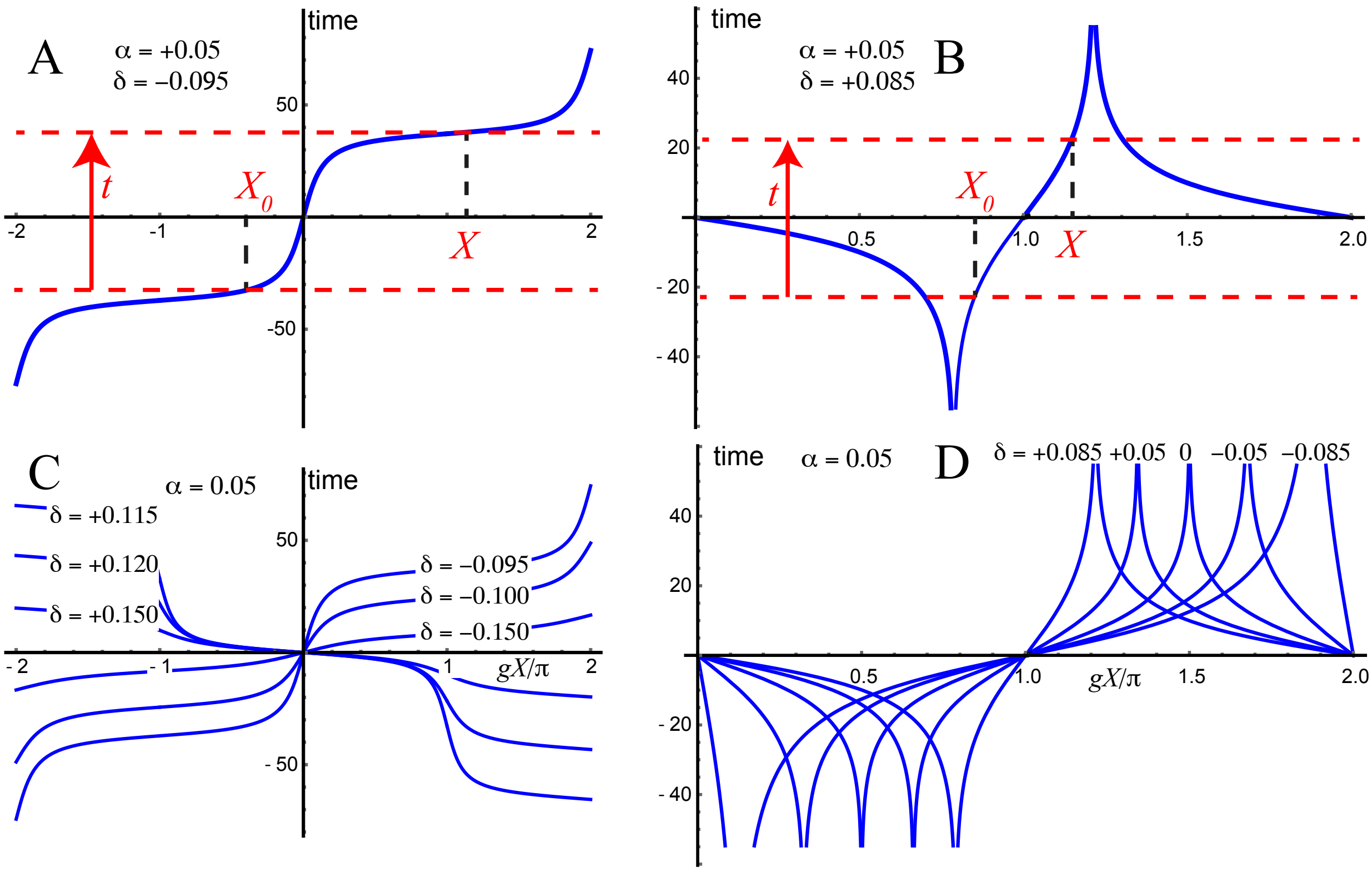

Fig. 2. The time trajectories $t(X)$ as functions of $X$. A and C: outside the critical velocity range, $|\delta/1+\delta| > |2\alpha|$, B and D: inside the critical points, $|\delta/1+\delta| < |2\alpha|$. The trajectories start at a generic $X_0$ where the value of $\psi$ is defined. This value, constant along the trajectory, travels to point $X$ after a time $t$. A and B illustrate the construction for tracing time back along a trajectory to the origin: the red lines show a simple construction for finding $X$. C and D show evolution of the trajectories with $\delta$ which defines the difference of the grating velocity of the average velocity of light. Note that the trajectories on the left are continuous and the ones on the right are singular, implying localisation.

We make use of (5) to calculate a dispersion relationship assuming a time dependence of $e^{-i\omega' t'}$ in the grating frame for the Bloch wave, and arrive at,

$$\frac{\partial}{\partial X}\ln\left[c_l - c_g\right] + \frac{\partial}{\partial X}\ln D_z = \frac{i\omega'}{c_l - c_g} \tag{12}$$

From which we deduce the change of phase across one unit cell which we equate to,

$$k_{eff}\frac{2\pi}{g} = \omega' \int_{X_0}^{X_0+2\pi/g} \frac{1}{c_l(X) - c_g} dX = \omega' t\left(X_0 + 2\pi/g\right) \tag{13}$$

where $k_{eff}$ is the Block wave vector and $t\left(X_0 + 2\pi/g\right)$ can be found from (11),

$$k_{eff} = -\frac{\omega + c_g k_{eff}}{c_g}\left[1 \pm \frac{1}{\sqrt{\delta^2 - 4\alpha^2(1+\delta)^2}}\right] \tag{14}$$

where we have substituting the frequency in the stationary frame. Solving for $k_{eff}$,

$$k_{eff} = \frac{\omega}{c_l} \frac{1 \pm \sqrt{\delta^2 - 4\alpha^2 (1+\delta)^2}}{1+\delta} \tag{15}$$

to give the effective velocity

$$c_{eff} = c_l \frac{1+\delta}{1 \pm \sqrt{\delta^2 - 4\alpha^2 (1+\delta)^2}} \tag{16}$$

where we choose the '+' sign when $\delta > 0$ and the '-' sign when $\delta < 0$. At the critical values of $\alpha$, $c_{eff}$ has a square root singularity as $c_{eff} \to c_g$ and the Bloch wave locks onto the grating. The variation of $c_{eff}$ with $\delta$ is shown in fig. 3. Note the shaded region within which Bloch waves are not defined, and the coincidence of $c_{eff}$ and $c_g$ at the boundary as the grating captures and localises the light. Below the lower cut-off $c_{eff} < c_\ell$ contrary to the predictions of Fresnel drag in a medium that is actually moving as observed in [11]. However above the upper cut-off $c_{eff} > c_\ell$. A better predictor of $c_{eff}$ in our example is that it is attracted to the velocity of the grating, increasingly so as the cut-off points are approached. Another interesting aspect is that $c_{eff}$, which serves as both the phase and group velocity of the Bloch wave, is superluminal when $c_g > c_\ell$. This happens only in systems operating a gain mechanism.

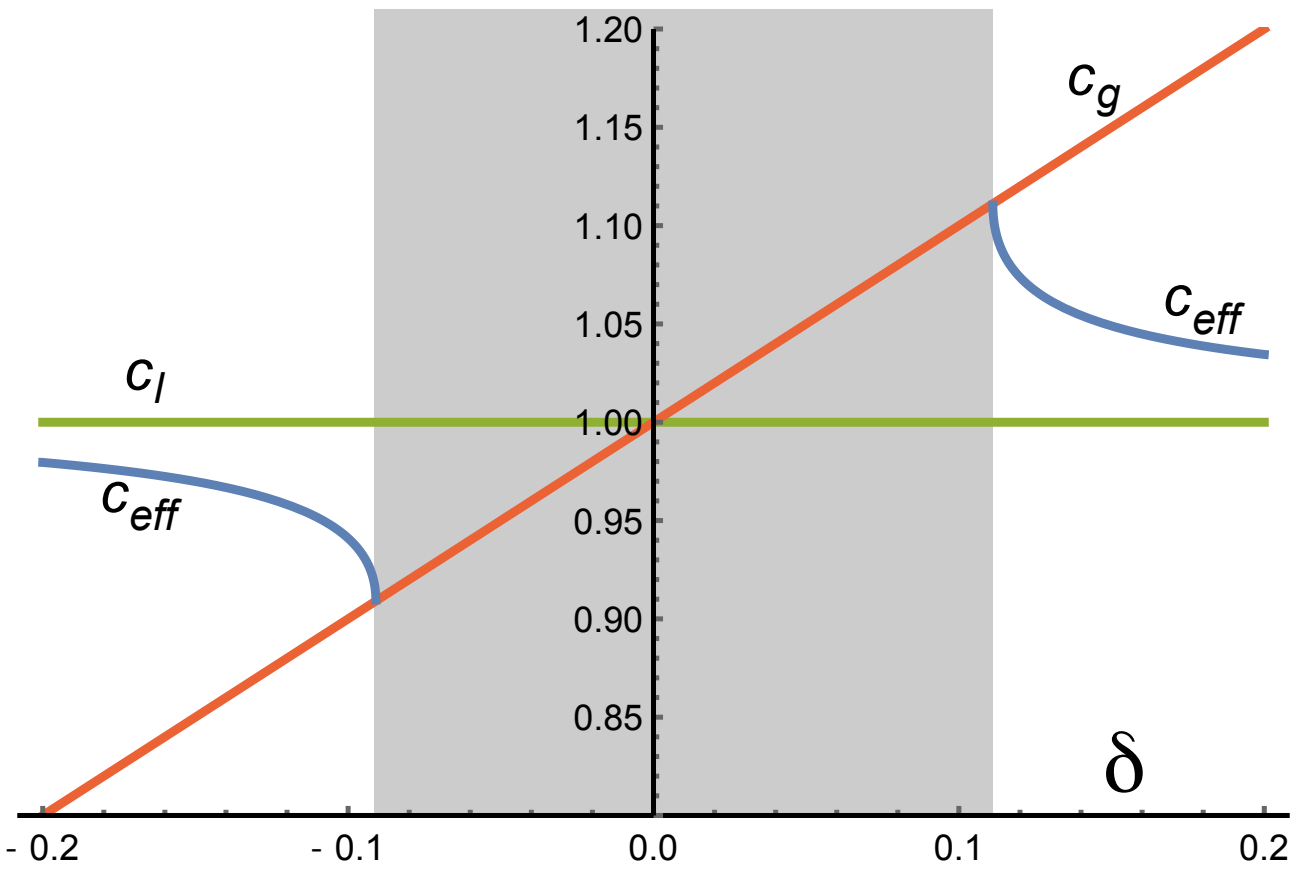

Fig. 3. The variation of $c_{eff}$ with $\delta$ for $\alpha = 0.05$, also showing the grating velocity $c_g = 1+\delta$ and the velocity of light in the background medium, $c_l$. Shading shows the range of $\delta$ within which Bloch waves are not defined.

However this simple effective medium picture hides a wealth of structure that we shall now investigate.

*The approach to criticality*

Using the characteristic trajectories we can easily find analytic solutions. There are two boundary conditions that could be imposed. In the first type the grating is turned on instantly everywhere; in the second the grating is turned on at a certain position in space and off at another. The latter is in fact a transmission calculation.

The boundary condition we have imposed requires the grating to be switched on at $t = t_0$, but if instead the light is entering a grating at some fixed point in space, the grating will be turned on at a time $t_0 - c_g^{-1} X_0$. and off at a time $t - c_g^{-1} X$. This small change in the boundary conditions means that (2) is now explicitly soluble,

First consider the case outside the critical region. Using the equations above we calculate,

$$gX_0 = -2\arctan \frac{\sqrt{\left(\frac{\delta}{1+\delta}\right)^2 - 4\alpha^2}}{\frac{\delta}{1+\delta} - 2\alpha} \tan\left[\frac{\theta(X)-t}{2} c_\ell g (1+\delta)^2 \sqrt{\left(\frac{\delta}{1+\delta}\right)^2 - 4\alpha^2}\right] \quad (17)$$

which is then used to calculate the fields,

$$D_z(X,t) = \frac{c_l(X_0(t)) - c_g}{c_l(X) - c_g} D_{z0} = \frac{1 + 2\alpha\cos(gX)}{\frac{\delta}{1+\delta} + 2\alpha\cos(gX)} \frac{\frac{\delta}{1+\delta} + 2\alpha\cos(gX_0(t))}{1 + 2\alpha\cos(gX_0(t))} D_{z0} \quad (18)$$

The ratio of numerator to denominator in this equation neatly expresses the compression of lines of force reported in an earlier paper [14]: the slower the velocity of waves relative to the grating the denser the lines of force, just as motorway traffic increases in density at slower speeds.

Evidently from (17) the solution is oscillatory with a period,

$$\Delta t = \frac{2\pi}{c_\ell g (1+\delta)^2 \sqrt{\left(\frac{\delta}{1+\delta}\right)^2 - 4\alpha^2}} \quad (19)$$

It has a maximum amplitude when $\delta < 0$ at $gX = \pi$, and when $\delta > 0$ at $gX = 0$,

$$\begin{aligned} D_{z,\max} &= \frac{1 + \frac{\delta}{1+\delta} + 2\alpha}{\frac{\delta}{1+\delta} + 2\alpha} \frac{\frac{\delta}{1+\delta} - 2\alpha}{1 + \frac{\delta}{1+\delta} - 2\alpha} D_{z0}, \quad \delta < 0, \\ &= \frac{1 + \frac{\delta}{1+\delta} - 2\alpha}{\frac{\delta}{1+\delta} - 2\alpha} \frac{\frac{\delta}{1+\delta} + 2\alpha}{1 + \frac{\delta}{1+\delta} + 2\alpha} D_{z0}, \quad \delta > 0 \end{aligned} \quad (20)$$

As the critical point is approached $\Delta t \to \infty$ with a critical exponent of $\upsilon = -1/2$ and $D_{z,\max} \to \infty$ with an exponent of $\tau = -1$.

These oscillations are displayed in fig. 4A for a value of $c_g$ above the critical velocity. Below in fig. 4B we plot $\varepsilon$. The value of $\varepsilon$ where $c_g = c_l$, shown notionally in the figure is not found anywhere within the grating but $c_l$ is closest to $c_g$ in the centre of fig. 4B. Therefore as the pulse moves to the left being overtaken by the grating it is first amplified slowing down as it moves until at $gX/\pi = 1.0$ it reaches its minimum velocity relative to the grating, after which it accelerates away, losing energy as it goes. In fig. 4C oscillations are shown for $c_g$ below the critical velocity. Here the pulse moves to the right as it overtakes the grating, gaining energy, and travelling ever more slowly until it reaches $gX/\pi = 0.0$ whereafter it accelerates away and loses energy. The variation of $\varepsilon$ is shown below in fig. 4D.

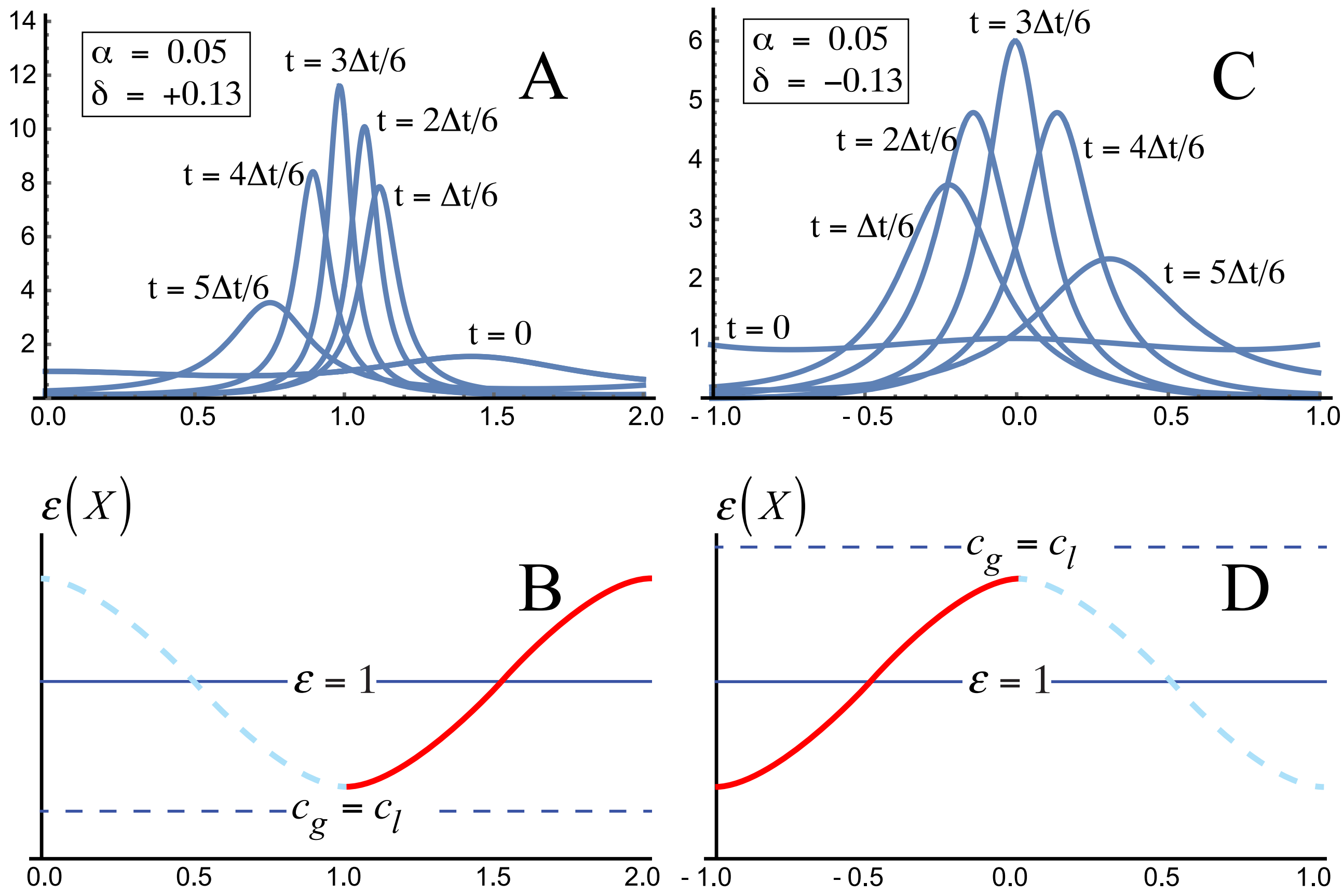

Fig. 4. Oscillations shown by A super and C sub luminal gratings plotted against $gX/\pi$ . The times are give as fractions of the period, $\Delta t$, as defined in (19). Below in B and D we show the variation of $\varepsilon$ across the grating and the dashed lines show schematically the values of $\varepsilon$ for which the grating and light velocities would be equal. The full line shows where gain occurs and dashed line shows where there is loss.

Our analytic theory has been verified by comparison with transfer matrix computations.

Next consider the case inside the critical region. Here we find,

$$\tan\left(gX_0/2\right) = +\frac{A}{B}\frac{-e^{-tc_\ell g(1+\delta)^2 a}\left(\frac{A}{B}-\tan\left(gX/2\right)\right)\left(\frac{A}{B}+\tan\left(gX/2\right)\right)^{-1}+1}{+e^{-tc_\ell g(1+\delta)^2 a}\left(\frac{A}{B}-\tan\left(gX/2\right)\right)\left(\frac{A}{B}+\tan\left(gX/2\right)\right)^{-1}+1}, \tag{21}$$

$$A = \sqrt{4\alpha^2 - \left(\frac{\delta}{1+\delta}\right)^2}, \quad B = 2\alpha - \frac{\delta}{1+\delta}$$

from which, using (17) we calculate the fields shown in fig. 5 for various values of the grating speed. In this paper we are interested in the approach to criticality: the upper critical velocity sees the maximum amplitude migrate to $gX = \pi$ and for the lower critical velocity to $gX = 2\pi$ . Here they meet up with the peak amplitudes outside the luminal region.

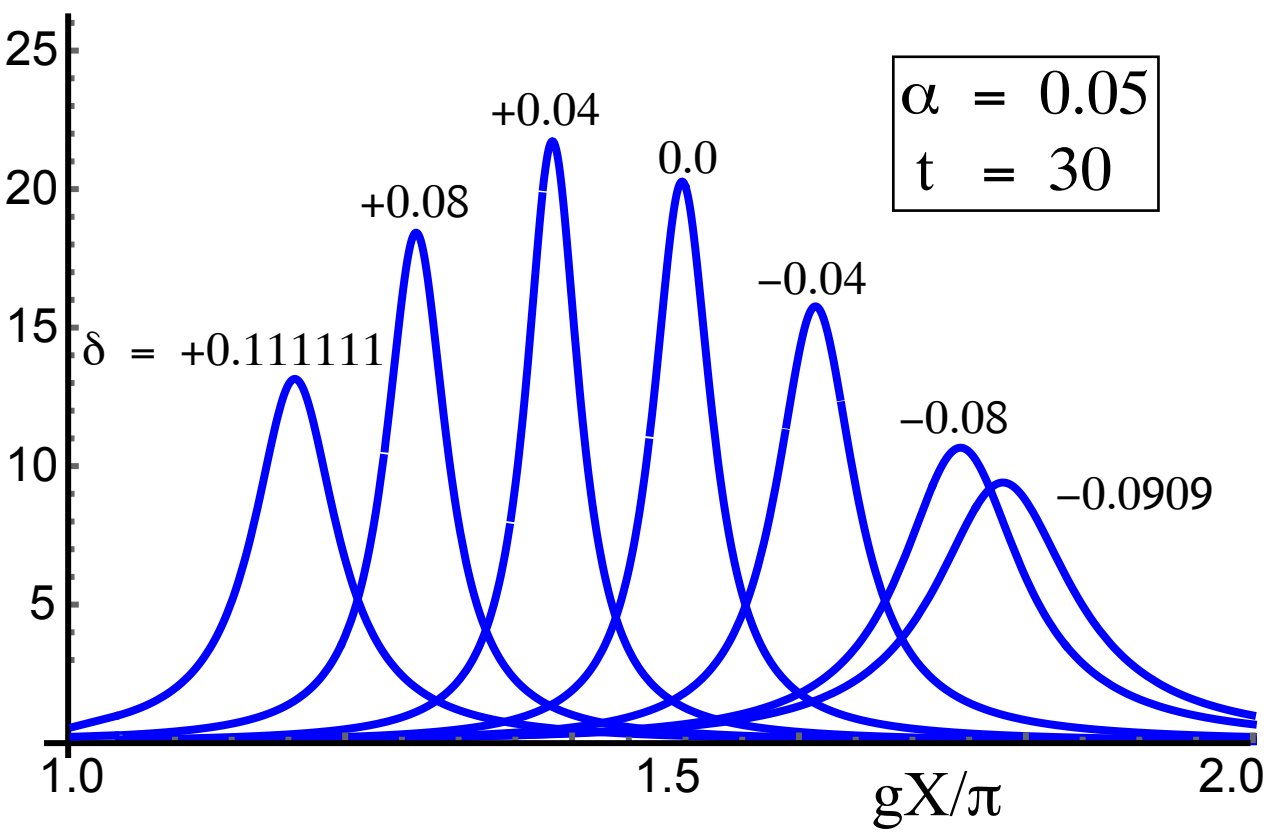

Fig. 5. Amplitude of the fields ejected from the latter half of the grating in the near luminal regime for various values of the grating speed, $c_g = c_1(1+\delta)$. The critical values for this example are $\delta = +0.111\cdots$, and $\delta = -0.0909\cdots$.

In the limit we have an analytic expression for the amplitude,

$$\lim_{\substack{t\to\infty \\ gX\to 0}} D_z(X,t) \approx \frac{\dfrac{\delta}{1+\delta} - 2\alpha}{\dfrac{\delta}{1+\delta} + 2\alpha - \alpha(gX)^2} D_{z0} \tag{22}$$

*Conclusions*

We have examined in detail the breaking of Bloch symmetry in a near luminal grating. Central to our work has been the newly applied "method of characteristics" which enables exact solution in a situation where forward scattering dominates. The latter comes about from the near degeneracy of forward travelling waves as $c_g \to c_l$. We observed oscillations in the transmission current in the Bloch regime with a period explained by the difference between $c_g$ and the Bloch group velocity, the wave being alternately amplified and attenuated as it rides over different sections of the unit cell. The critical exponent of $1/2$ with which the relative velocity of the wave and the grating vanishes at the transition point is reflected on the luminal side of the transition in the same critical exponent with which the exponential increase of amplitude growth initiates. In summary: a detailed understanding has been found of the symmetry breaking transition observed in luminal structures.

*Acknowledgments*

P.A.H. and M.S. acknowledge funding from Fundação para a Ciencia e a Tecnologia and Instituto de Telecomunicações under Project UID/EEA/50008/2020. P.A.H. is supported by the CEEC Individual program from Fundação para a Ciencia e a Tecnologia and Instituto de Telecomunicações with reference CEECIND/03866/2017. M.S. acknowledges support from the IET under the A F Harvey Prize and by the Simons Foundation under the award 733700 (Simons Collaboration in Mathematics and Physics, “Harnessing Universal Symmetry Concepts for Extreme Wave Phenomena”). E.G. acknowledges support from the Engineering and Physical Sciences Research Council through the Centre for Doctoral Training in Theory and Simulation of Materials (grant EP/L015579/1) and an EPSRC Doctoral Prize Fellowship (grant EP/T51780X/1). J.B.P. acknowledges funding from the Gordon and Betty Moore Foundation.